\documentclass[12pt,titlepage,biophysj]{article}

\newcommand{\B}{\mathcal{B}}
\newcommand{\LL}{\mathcal{L}}
\newcommand{\MM}{\mathcal{M}}
\newcommand{\HH}{\mathcal{H}}
\newcommand{\1}{$^{-1}$}
\newcommand{\dg}{$^{\circ}$}
\newcommand{\bea}{\begin{eqnarray}}
\newcommand{\eea}{\end{eqnarray}}
\newcommand{\be}{\begin{equation}}
\newcommand{\ee}{\end{equation}}
\newcommand{\kB}{k_{\rm B}}
\usepackage{graphicx}
\usepackage{color}
\usepackage{amsmath}
\usepackage[latin1]{inputenc}
\usepackage[tight]{subfigure}
\usepackage[round,numbers,sort&compress]{natbib} 
\begin{document}

\title{Stepwise bending of DNA by a single TATA-box Binding Protein}

\author{Simon F. Toli{\'{c}}-N\o{}rrelykke*$^{\#\S}$\thanks{
Corresponding author.
Address: Max-Planck Institute for the Physics of Complex Systems, N\"othnitzerstr.\ 38, 01187 Dresden, Germany,
Tel.: (+49) 351 871 1211, Fax: (+49) 351 871 1999, E-mail: simonftn@pks.mpg.de}\\
\and Mette B. Rasmussen*\\
\and Francesco S. Pavone$^{\#}$\\
\and Kirstine Berg-S\o{}rensen*$^{\P}$\\
\and Lene B. Oddershede*\\
\and \\ 
\small *The Niels Bohr Institute, DK-2100 Copenhagen \O, Denmark\\
\small $^{\#}$European Laboratory for Non-linear Spectroscopy, 50019 Sesto Fiorentino (Fl), Italy\\
\small $^{\S}$Max-Planck Institute for the Physics of Complex Systems, 01187 Dresden, Germany\\  
\small $^{\P}$Present address: Dept.\ of Physics, Technical University of Denmark\\ 
\small DK-2800, Kgs.\ Lyngby, Denmark\\}

\date{\today}
\pagestyle{myheadings}
\markright{Stepwise TBP-DNA interaction}

\maketitle

%%%%%%%%%%%%%%%%%%%%%%%%%%%%%%%%%%%%%%%%%%%%%%
\abstract{
The TATA-box Binding Protein (TBP) is required by all three eukaryotic RNA polymerases for the initiation of transcription from most promoters.  
TBP recognizes, binds to, and bends promoter sequences called ``TATA-boxes'' in the DNA\@. 
We present results from the study of individual \textit{Saccharomyces cerevisiae} TBPs 
interacting with single DNA molecules containing a TATA-box.
Using video microscopy, we observed the Brownian motion of beads tethered by  
short surface-bound DNA\@. 
When TBP binds to and bends the DNA, the conformation of the DNA changes and the amplitude of Brownian motion of the tethered bead is reduced compared to that of unbent DNA\@.
We detected individual binding and dissociation events and derived kinetic parameters for the process.
Dissociation was induced by increasing the salt concentration or by directly pulling on the tethered bead using optical tweezers.
In addition to the well-defined free and bound classes of Brownian motion, we observed another two classes of motion.
These extra classes were identified with intermediate states on a three-step, linear binding  pathway.
Biological implications of the intermediate states are discussed.}

\vspace{1cm}
\clearpage
%\twocolumn

%%%%%%%%%%%%%%%%%%%%%%%%%%%%%%%%%%%%%%%%%%%%%%%

\section*{INTRODUCTION}
Observation of  biological systems on the single-molecule level can reveal information that is not easily obtainable in ensemble studies.
By studying individual molecules it is possible to address whether an observed variation in activity is caused by temporal variations in the individual molecules or by variation in activity from molecule to molecule \cite{Lu1998,Neuman2003,Tolic-Norrelykke2004}.
Also, single-molecule techniques often allow the application of a mechanical force to the system and thus the introduction of a well defined reaction coordinate \cite{Bustamante2000}.

The TATA-box Binding Protein is a small, single-chain, saddle-shaped protein \cite{Nikolov1992}, the DNA-binding part of which is highly conserved throughout evolution.
The inner, concave side of TBP directly contacts the DNA, whereas the outer and evolutionary more variable side interacts with various other proteins involved in the regulation of transcription \cite{Alberts2002}.
With TBP bound, the DNA is bent by $\sim$80$^{\circ}$ and locally unwound by $\sim$120$^{\circ}$ \cite{Kim1993admlp,Kim1993cyc1}.
DNA distortion is thought to play a role in transcription initiation, the distortion influencing recruitment and stabilization of RNA polymerases and associated proteins.
Furthermore, this distortion may have a direct mechanical role in chromatin remodeling: 
Histones have been shown to slide along the DNA upon TBP binding \cite{Lomvardas2001}.
Association of TBP with promoter DNA is a slow process, but after binding the complex can support multiple rounds of transcription initiation \cite{Yean1997}.

Most DNA-binding proteins interact with DNA through the major groove, where the base-paired sequence is easily accessible.
TBP, however, interacts with DNA through the minor groove \cite{Kim1993admlp,Kim1993cyc1} in a manner that differs also from that of other minor-groove-binding proteins \cite{Kim1994}.
The TBP-DNA interaction is well studied at the ensemble level:
Bending angles \cite{Kim1993cyc1,Kim1993admlp,Wu2001b} as well as the specificity of the binding \cite{Hahn1989,Hoopes1992,Perez-Howard1995,Parkhurst1996,Parkhurst1999,Wu2001a,Wu2001b} to different DNA sequences under various conditions of pH, temperature, osmolyte, and electrolyte concentrations have been revealed using X-rays, electron microscopy, gel-retardation, DNase I foot-printing, fluorescence anisotropy, and F\"orster Resonance Energy Transfer (FRET). 
However, no previously reported studies have addressed this system at the single-molecule level.

TBP is known to bind to several consensus and non-consensus TATA-boxes \cite{Hahn1989}.
The best studied TATA-box is the adenovirus major late promoter (AdMLP), 
which serves as a reference example of TBP-DNA interactions.
When binding to the AdMLP, TBP has to overcome an activation barrier of nearly 
10 kcal/mol, but once bound, the protein resides in an energetic minimum 
that is almost 11 kcal/mol deep \cite{Parkhurst1999}.
A careful analysis of ensemble data taken at a range of temperatures and protein concentrations led to a prediction of two intermediate states on the reaction pathway between the initial, unbound state and the final, bound state \cite{Parkhurst1999}.  
The structure of the TBP-DNA complex in these intermediate states is not known, but was proposed \cite{Parkhurst1999} to be nearly identical to the final bound form.
However, direct observation of these intermediates has not previously been reported.

We set up a tethered-particle-motion system \cite{Schaefer1991,Yin1994} to study the interaction between TBP and DNA at the single-molecule level.  
A microsphere (``bead'') was tethered to a microscope coverslip by 324\,bp of DNA\@. 
In the center of the DNA sequence we placed a TATA-box, the rest of the DNA contained no TATA-like sequences. 
Using video microscopy we followed the Brownian motion of the bead and from this motion constructed a measure for the conformation of the DNA tether.
Upon introduction of TBP, we observed a decrease in Brownian motion and interpreted this decrease as the binding of TBP to the DNA\@.
A series of control experiments were undertaken to ensure that the observed effect was caused by active TBP, binding specifically to the TATA-sequence.
We showed that TBP can be forced off the TATA-box sequence by increasing the electrolyte concentration or by mechanically pulling on the DNA using laser tweezers.
The choice of a short DNA-tether and relatively large beads ensured that any change in conformation of the DNA was amplified into a larger change in the position of the tethered bead.  
Using this system, we studied the binding kinetics of TBP to DNA and showed by direct observation that at least one of the intermediate states is less bent than the final state.
This constitutes the first direct observation of a structural intermediate on the TBP-DNA binding pathway, a finding that may have implications for our understanding of the regulation of transcription initiation.

%%%%%%%%%%%%%%%%%%%%%%%%%%%%%%%%%%%%%%%%%%%%%%%
\section*{MATERIALS AND METHODS}
\subsection*{DNA}
DNA-tethers for binding experiments (``TATA-DNA'') were engineered to have the AdMLP sequence 5'-TATAAAAG-3' at position 155.
324-bp double stranded DNA labeled with digoxigenin at one end and biotin at the other was produced by 10 rounds of polymerase chain reaction (PCR) amplification from pSF1 (see \cite{Simon_phd} for details).
DNA-tethers without the TATA-box (``control-DNA'') were produced by site-directed mutagenesis using PCR to have the \textit{Xho}I recognitions sequence (underlined) 5'-TA\underline{CTCG}AG-3' at position 155.
We screened the DNA for consensus as well as non-consensus TBP binding sequences 
known from the literature \cite{Singer1990,Starr1995,Wobbe1990,Wu2001a,Powell2001,Hoopes1998}. 
In the TATA-DNA we found no high-affinity sequences other than the AdMLP TATA-box and in the control-DNA we found no high-affinity sequences.
%%%%%%%%%%%%%%%%%%%%%%%%%%%%%%%%%%%%%%%%%

\subsection*{Beads}
\begin{figure}[h] 
\begin{center}
	\includegraphics*[width= \linewidth]{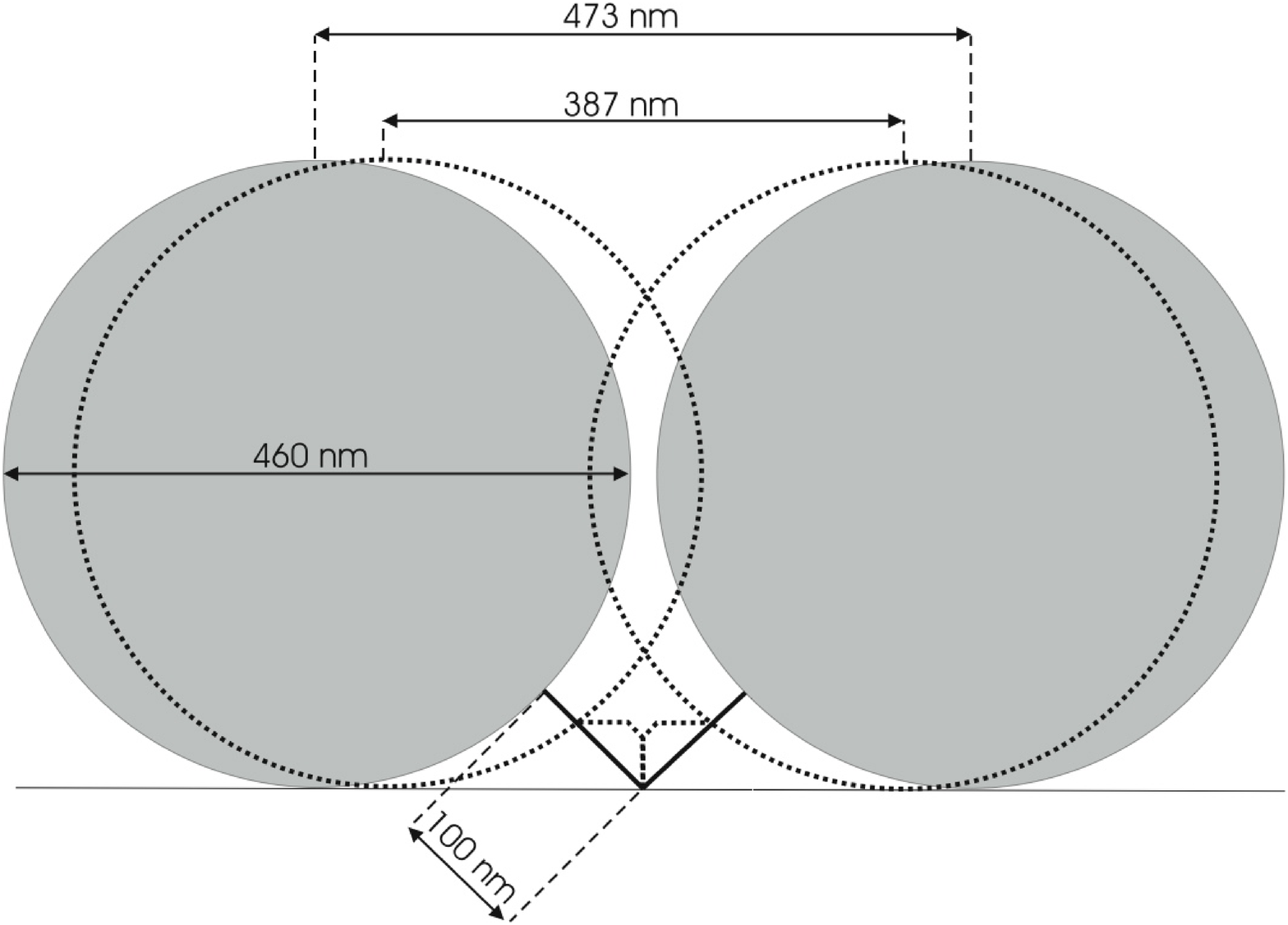}
	\caption{}
	\label{Fig:scaledraw}
\end{center}
\end{figure}

The beads used were streptavidin coated 
polystyrene spheres of diameter $d$\,=\,0.46\,$\mu$m (BangsLabs),
as well as 1.0\,$\mu$m streptavidin coated silica spheres 
(Spherotech).
One end of the DNA tether was specifically attached to
the avidin coated micro-sphere by biotin-streptavidin binding. 
The other end was attached to a glass coverslip by digoxigenin/anti-digoxigenin
interaction. 
Nuclease-free bovine serum albumin (BSA; Roche Applied Sciences) was added to the beads to a final concentration of 1\,mg/ml and incubated for $>$14\,h at 4\,\dg C before use, to suppress sticking to the glass coverslip.
Figure \ref{Fig:scaledraw} shows a schematic drawing of the setup.

\subsection*{TATA-box Binding Protein}
\textit{Saccharomyces cerevisiae} TBP was prepared as described in \cite{Parkhurst1996,Petri1995}.
The protein was stored at -80$^{\circ}$C, at a concentration of 33.9 $\mu$M, in Buffer\,A
(25\,mM Hepes-KOH, 2.7\,M glycerol, 1\,mM EDTA, 1\,mM DTT, pH 7.9)
until immediately before use, at which time it was thawed, diluted in Buffer 1
(10\,mM Tris-HCl, 100\,mM KCl, 2.5\,mM MgCl$_2$, 1\,mM CaCl$_2$, 1\,mM DTT, pH 7.4)
to the working concentration of 50--678\,nM, and kept on ice between experiments.
The stock concentration of TBP was determined from the absorbance at 280\,nm using an extinction coefficient of 12,700\,M$^{-1}$cm$^{-1}$ \cite{Petri1995}.
TBP is monomeric under the conditions and concentrations used in the experiments described here \cite{Parkhurst1999,Daugherty1999}.

\subsection*{Preparation of Samples}
Microscopy flow cells ($\sim$20\,$\mu$l) were prepared by placing one coverslip No.~0 on top of another, separated by spacers made from a single thickness of Parafilm (American National Can, Menasha, WI, USA) coated with silicone vacuum grease (Dow-Corning).
Anti-digoxigenin was attached to the working surface by injecting the flow cell
with 20\,$\mu$g/ml anti-digoxigenin in PBS and incubating at 4$^{\circ}$C for $>$14\,hours. 
The flow cells were then washed with 3$\times$100\,$\mu$l Buffer\,1, 
incubated with DNA for 15\,min, and again washed with 3$\times$100\,$\mu$l Buffer\,1.
To suppress non-specific binding of beads and TBP to the surface, the flow cells were incubated with 2\,mg/ml BSA in Buffer\,1 for 15\,min saturating all surfaces with BSA\@.
Samples were mounted on a microscope for observation, and beads were flowed in at high concentrations (40--160\,pM) and allowed to form tethers for 20\,min before unbound beads were removed by washing with 3--5$\times$100\,$\mu$l Buffer\,1.

%%%%%%%%%%%%%%%%%%%%%%%%%%%%%%%%%%%%%%%%%%%%%%

\subsection*{Video Microscopy and Optical Tweezers}
Time-lapse video images for the experiments using silica beads were acquired using a modified microscope (Leica DM IRB, 100$\times$ oil-immersion objective, \cite{Oddershede2001}).
Bead positions were determined in each video-frame using a cross-correlation type tracking algorithm written in MatLab (The MathWorks, MA, USA) \cite{Simon_phd}.
The laser tweezers experiments were also performed on this setup.

To zoom in (temporally) on the binding step, a different microscope was used (Nikon Eclipse TE300,  60$\times$ water immersion objective, \cite{Capitanio2002}).
Bead positions were determined by using a thresholded centroid-tracking method for each of the two de-interlaced video-fields that comprise a video-frame, thus obtaining a time resolution of 1/50\,s.

%%%%%%%%%%%%%%%%%%%%%%%%%%%%%%%%%%%%%%%%%%%

\subsection*{Brownian Motion Measure}
The shortness of the DNA ensured that tethered beads stayed in the focal plane of the microscope.
Motion perpendicular to the surface was not detected.
Motion parallel to the surface, in the ($x,y$)-plane, was detected using one of the two 
above mentioned tracking algorithms.
Time-series of $(x,y)$-positions were broken into non-overlapping segments and the variances, $\sigma_x^2$ and $\sigma_y^2$, were calculated in each segment.
As measure for the Brownian motion of a tethered bead, we chose the root-mean-square-deviation (RMSD), $\B=\sqrt{(\sigma_x^2 + \sigma_y^2)/2}$.
Additional measures were applied that allowed us to distinguish between beads tethered by one, two, or more DNA-molecules (see Appendix~\ref{app:RMSD} and \ref{app:tethers}, and Fig.~\ref{Fig:double_tether_cut}).
%%%%%%%%%%%%%%%%%%%%%%%%%%%%%%%%%%%%%%%%%%%%%%%
\subsection*{Establishment of Method}
To investigate the nature of the tether between the bead and the coverslip,
and the effect of TBP, we did a series of control experiments: 
\subsubsection*{Tethers}
A concentration series of DNA showed that the number of tethered beads
varied linearly with the amount of DNA, with few or no beads tethered when no DNA was present and saturation at high DNA concentrations. 
Flow cells prepared without anti-digoxigenin lead to a $>$90\% decrease in the number of tethers formed, as did  pre-blocking the streptavidin coated beads with saturating amounts of biotin. 

\subsubsection*{Effect of glycerol}
The presence of osmolytes, e.g.\ glycerol and glucose, have been reported to dramatically increase both binding affinity and bending angle for TBP binding to DNA \cite{Wu2001b}.
Since our protein was stored in a glycerol-rich buffer, we decided to investigate the effect of glycerol on our single-molecule experiments.
We added 226\,nM TBP in 0.91\,M glycerol (1/3 Buffer\,A and 2/3 Buffer\,1) to flow cells with beads tethered by either TATA-DNA or control-DNA\@.
Essentially all of the tethered beads showed large and instant decreases in Brownian RMSD, making it impossible to distinguish between beads tethered by TATA-DNA and beads tethered by control-DNA\@.
To avoid this effect, all experiments reported hereafter were conducted at 4--18\,mM glycerol.
This is within the range of intracellular concentrations found in \textit{S. cerevesiae} at different growth stages \cite{Cronwright2002}, and below the concentration where bend-angles are expected to be affected \cite{Wu2001b}.
The effect of osmolytes on the binding to nonspecific versus target DNA was recently reported in detail for the \textit{lac} repressor system \cite{Fried2002}.

\subsubsection*{Protein}
TBP not competent in binding DNA might interact with the bead directly by
sticking to the glass and the bead simultaneously, or indirectly through entropic effects \cite{Crocker1999}.
Such interaction would influence the motion of the bead and hence also the conformations of the DNA\@.
To clarify whether any such effects were present, we conducted experiments with heat-inactivated protein:
TBP was incubated at 100$^\circ$C for 5\,min, allowed to cool to room temperature, 
and then added at a concentration of 226\,nM to microscopy samples 
with either TATA-DNA tethers or control-DNA tethers.
No change in the Brownian RMSD of the tethered beads was observed in samples where inactivated TBP was added.

\subsubsection*{Specificity of TBP binding}
\begin{figure}[h] 
\begin{center}
	\includegraphics*[width= \linewidth]{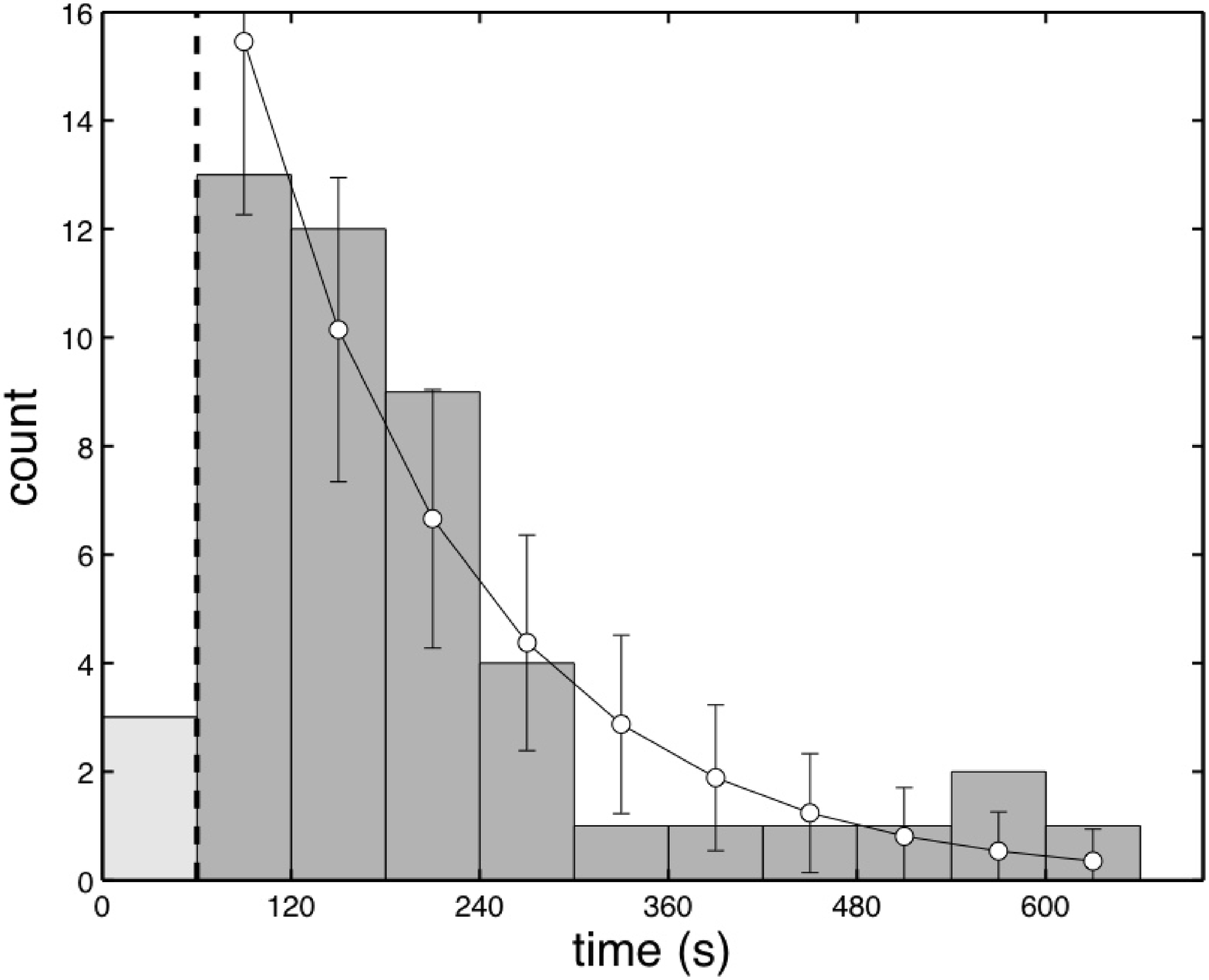}
	\caption{}
	\label{Fig:kinetics}
\end{center}
\end{figure}

\begin{figure}[h] 
\begin{center}
	\includegraphics*[width= \linewidth]{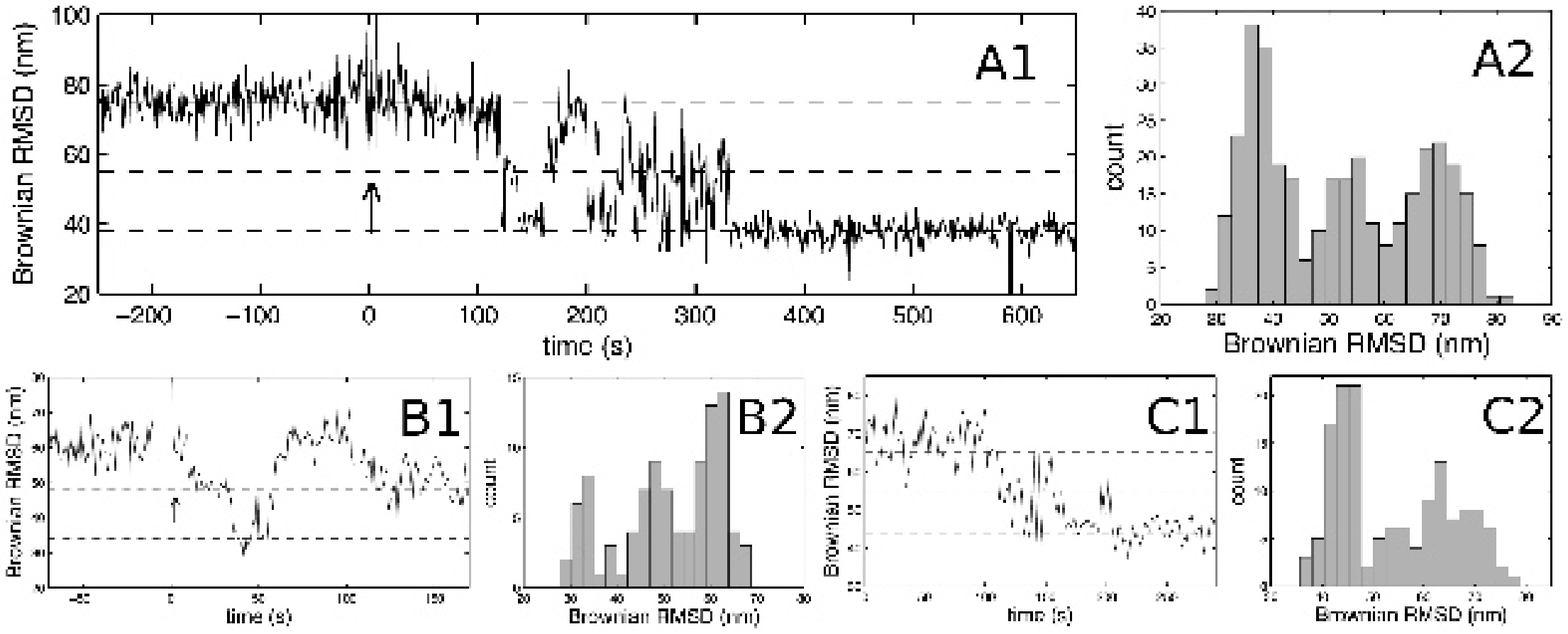}
	\caption{}
	 \label{Fig:steps_twocol}
\end{center}
\end{figure}

To investigate whether the interaction between TBP and DNA
was confined to the TATA-box of the DNA we used the control-DNA described above. 
Any effect of TBP on the Brownian motion of beads tethered by control-DNA would be caused by non-specific interactions between TBP and DNA, since no TATA-box is present in this tether.
At low concentrations of TBP ($<$100\,nM) we detected no change in the Brownian RMSD of beads tethered by control-DNA\@.
At intermediate concentrations (226\,nM) a small fraction of beads tethered by control-DNA showed brief ($\sim$1\,min), minor ($\sim$10\,nm) decreases in Brownian RMSD\@.
At high concentrations of TBP (678\,nM) beads tethered by control-DNA showed abrupt
decreases in Brownian RMSD, often resulting in the bead becoming irreversibly stuck on the coverslip---this behavior was also observed for beads tethered by TATA-DNA\@.
We interpret this latter high-concentration-behavior as the ``collapse'' of DNA induced by non-specific action of the protein, a process also observed in the non-specific interaction of other proteins with DNA \cite{Ali2001,Brewer2003}.
To collect the data summarized in Fig.~\ref{Fig:kinetics} we used 226\,nM TBP, and for the data shown in Fig.~\ref{Fig:steps_twocol} we worked at 50--100\,nM TBP\@.

%%%%%%%%%%%%%%%%%%%%%%%%%%%%%%%%%%%%%%%%%%%%%%%%%%%%%%%%%%%%%%%%%%%%%%%%%%%%%%%%%%%%%%%%%%%%%%%%%%%%
\section*{RESULTS AND DISCUSSION}

\subsection*{Binding and Dissociation Events}
\begin{figure}[h] 
\begin{center}
	\includegraphics*[width=\linewidth]{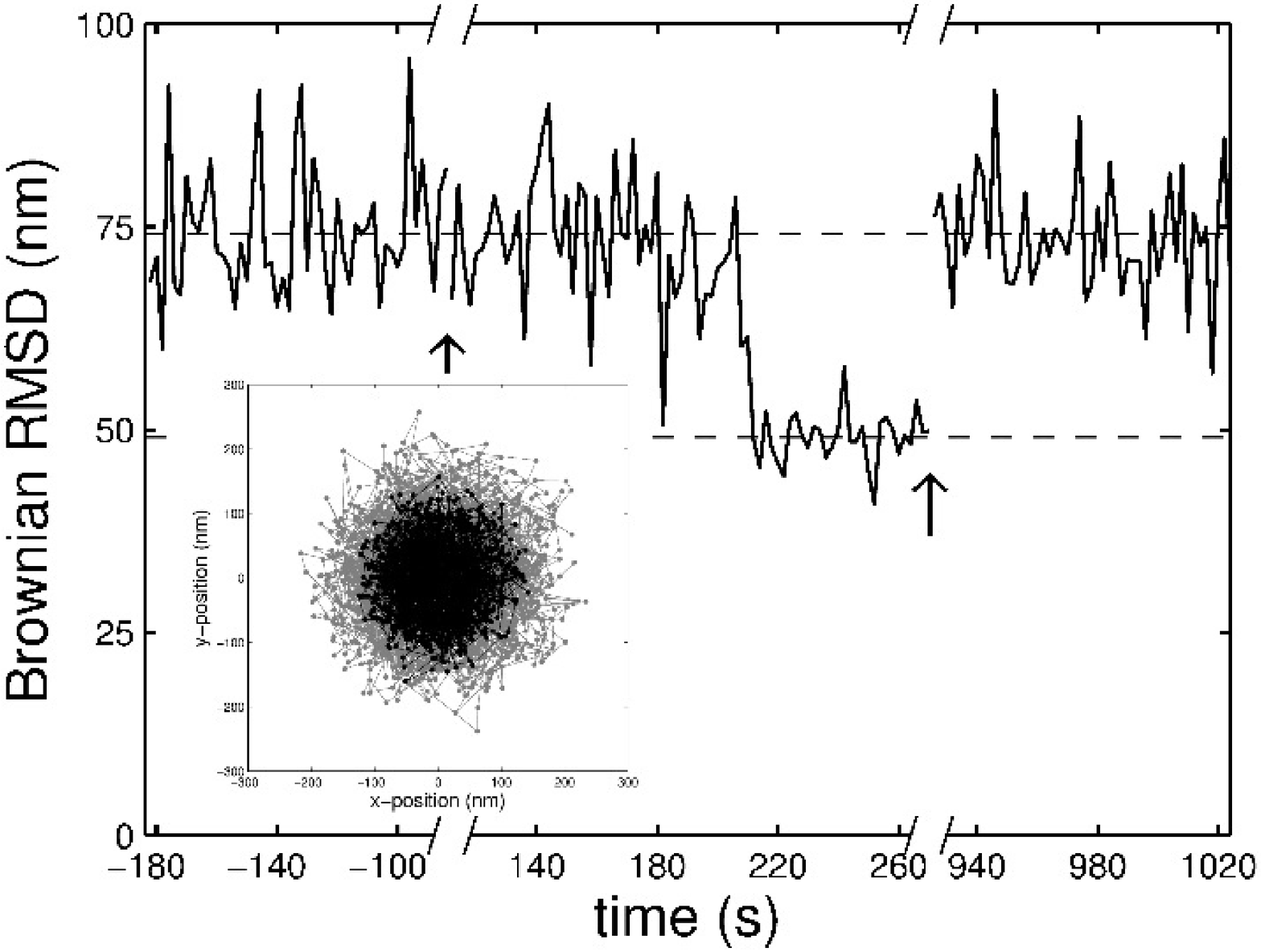}
	\caption{}
	\label{Fig:silica_event}
\end{center}
\end{figure}

Figure~\ref{Fig:silica_event} shows the Brownian RMSD, as a function of time, for a single DNA-tether containing a TATA-box:
Initially (negative times) no TBP was present and $\B$\,$\simeq$\,75\,nm; 
200\,s after addition of TBP, $\B$ decreased abruptly to $\sim$50\,nm; finally, after exposure to 1\,M KCl and washing with Buffer~1, the Brownian RMSD returned to 75\,nm.
We interpret this type of behavior as the specific binding of TBP to the recognition sequence in the middle of the DNA, concurrent with a decrease in Brownian RMSD (henceforth referred to as a ``binding event''). 
Addition of a high concentration of salt was followed by an increase in Brownian RMSD\@. We interpret this increase as the release of TBP from the TATA-box (henceforth referred to  as a  ``dissociation event'').
The inset in Fig.~\ref{Fig:silica_event} shows a scatterplot of the positions visited by the DNA-tethered bead. 
The recorded positions visited before TBP bends the DNA-tether are shown as grey points connected by lines. 
Black dots and lines show the recorded positions visited after TBP has bound to and bent the DNA-tether.
The positions visited by the bead are isotropically distributed in the $(x,y)$-plane, both before and after addition of TBP\@.

The Brownian RMSD, both before and after addition of TBP, varied from one tether to the next. 
In $n$\,=\,48 observed binding events using DNA-tethered silica beads the Brownian RMSD decreased from 69\,$\pm$\,13\,nm to 46\,$\pm$\,12\,nm (mean $\pm$ SD).
That is, the step-size, the change in Brownian RMSD upon binding by TBP, was $25 \pm 9$\,nm.

When flowing in 1\,M KCl we consistently, after a brief waiting time, saw $\B$ return to the value it had prior to the addition of TBP\@. 
We interpret this as the release of individual molecules of TBP from single DNA-tethers.  
Any non-specific, as well as specific, interaction keeping the TBP in contact with the DNA under normal conditions is suppressed at this concentration of salt because the electrostatic interactions are screened.  
Thus, once no longer bound to the TATA-box, the protein is expected to diffuse away from the DNA-tether.

Another way to force TBP off DNA is by stretching the
tether using an externally applied mechanical force \cite{Marko1997}. 
As a proof of principle, experiments were performed using laser tweezers to pull on
the bead, thereby stretching the DNA and forcing the protein off.
These experiments are described in Appendix~\ref{app:laser}. 
Our observations are in accordance with previously published results, showing that
proteins do not stay as readily attached to stretched DNA as to relaxed DNA \cite{Harada1999,Dixit2005}.

\subsubsection*{Binding kinetics}
We investigated the binding kinetics by measuring the time elapsed between the addition of TBP and the first detected decrease in Brownian RMSD\@.
In Fig.~\ref{Fig:kinetics}, a histogram of these waiting times is shown.
If the binding of protein to DNA has a constant probability of success per time interval it is a Poisson process.
The waiting times are then exponentially distributed.
A fit of a single exponential to the data returned a time-constant of 143\,s.
The corresponding second-order association rate constant is $k_a = \rm{[TBP]}^{-1} \tau^{-1} = (3.1 \pm 0.5) \times 10^4$\,M$^{-1}$s$^{-1}$ (mean $\pm$ SE, $n=45$).
This value is in rough agreement with values determined in ensemble experiments performed under similar buffer conditions and temperature  ($k_a = 8$--$14 \times10^{4}$\,M\1s\1 \cite{Parkhurst1999,Powell2001}).
The difference, if any,  could be caused by a lower than assumed concentration of binding-competent TBP in our experiments---the presence of surfaces is expected to diminish the amount of TBP through adsorption and denaturation \cite{Edwards1992,Craig1996}.
Furthermore, the protein is very sensitive to repeated freeze/thaw cycles:
We kept these to a minimum, but the experimental protocol meant that a few cycles of freeze/thaw could not be avoided.
Finally, the proximity of the tethered bead to the coverslip gives rise to a force on the DNA-tether due to excluded volume effects \cite{Segall2005}.
This force is expected to suppress the rate of association by 25--50\% (see Appendix~\ref{app:entropy}).

 %%%%%%%%%%%%%%%%%%%%%%%%%%%%%%%%%%%%%%%%%%%%%%

\subsubsection*{Dissociation kinetics}
Characteristic times for the single-exponential decay of the TBP-AdMLP complex have been reported in the range from 10 to 170\,min \cite{Parkhurst1999,Perez-Howard1995,Hoopes1998,Hoopes1992,Kuddus1993}.
The existence of an additional, fast phase in the dissociation, nearly two orders of magnitude faster than the dominant slow phase,
was revealed only by detailed kinetic studies \cite{Parkhurst1999}.

We did not determine the kinetics of spontaneous dissociation in our experiments.
An experiment was typically ended after 15--20\,min by exchanging the assay-buffer with 1M\,KCl, and then looking for dissociation events.
We did this to confirm that any observed change in Brownian RMSD was caused by TBP-DNA interactions. 
Furthermore, tethered beads had a tendency to get stuck on the coverslip after extended observation, thus terminating any further observation.

%%%%%%%%%%%%%%%%%%%%%%%%%%%%%%%%%%%%%%%%%%%%%%%
\subsection*{Intermediate States} 
To further investigate the binding kinetics, we modified the experimental setup.
By using polystyrene beads of a smaller diameter (0.46\,$\mu$m instead of 1.0\,$\mu$m) as well as separately analyzing each of the two video-fields that constitute a video-frame, the temporal resolution was increased.
The TBP concentration was 50--100\,nM, i.e., lower than in the experiments described above.  At these protein concentrations approximately 50\% of the tethers showed binding events.
The expected equilibrium dissociation constant for the TBP-DNA interaction is 16\,nM, with the buffer and temperature conditions used here \cite{Parkhurst1999}.  
However, due to the previously mentioned effects of surfaces, freeze/thaw cycles, and tension in the DNA, the observed single-molecule dissociation constant is expected to be higher.
During 10 out of 11 observed binding events, the Brownian RMSD was observed to decrease and sometimes increase in a stepwise manner in the presence of TBP, see Fig.~\ref{Fig:steps_twocol}.
The kinetic scheme suggested in \cite{Parkhurst1999} (boxes added by us):
\begin{displaymath}
 \underset{\HH}{\fbox{\rule[-0.35cm]{0cm}{0.9cm} DNA + TBP}} 
\overset{k_1}{\underset{k_2}{\rightleftharpoons}} \underset{\MM}{\fbox{\rule[-0.35cm]{0cm}{0.9cm} $I_1$}}
\overset{k_3}{\underset{k_4}{\rightleftharpoons}} \underset{\LL}{\fbox{ $I_2
\overset{k_5}{\underset{k_6}{\rightleftharpoons}} \mbox{DNA:TBP}_{\rm final}$}}
\enspace,
\end{displaymath}
predicts the existence of two intermediate states $I_1$ and $I_2$, on the path to the final, bound state. 
The intermediates and the final complex were all postulated to have the same bend in the DNA and to differ only in their stability \cite{Parkhurst1999}.

In an experiment where only the first and the last state is resolved the forward reaction can be described by a second-order association rate constant $k_a$ that depends on the microscopic rate constants $k_i$, $i=1$\dots6 \cite{Parkhurst1999}.
This rate-constant $k_a$ is the one cited above, because we were not able to discern intermediates using the silica beads (chosen for their high optical trapping efficiency, see Appendix~\ref{app:laser}) 
In the following we distinguish between the observed \textit{classes} of Brownian motion, and their interpretation in terms of different underlying \textit{states} of the TBP-DNA complex. 

\subsubsection*{Three classes of Brownian motion were observed in the histograms}Figure~\ref{Fig:steps_twocol} shows three examples of the time-development of $\B$ as well as histograms of $\B$.
Each histogram shows a distribution of the Brownian RMSD with three separate peaks.
Based on such histograms we divided the Brownian motion into three classes.
Each of these three classes must correspond to a different conformation of the TBP-DNA complex:
i) The $\HH$-class is defined from the Brownian RMSD observed before addition of TBP\@. 
Thus, the $\HH$-class corresponds to DNA that is not bound by TBP, but includes also the transient encounter complex that is implicit in all bimolecular reactions.
ii) The $\LL$-class is defined from the Brownian RMSD observed several minutes after addition of TBP\@.
We identify this class with a superposition of the final bound state and the second intermediate state $I_2$ (see discussion below).
iii) The $\MM$-class was a surprise.  
It is defined from the Brownian RMSD that was of a magnitude midway between the $\HH$ and $\LL$ classes.
We interpret this class as corresponding to the first intermediate state $I_1$.

Each of these three classes  is indicated by a box in the kinetic scheme shown above. 
The observed Brownian RMSD in each class was:
71 $\pm$ 7\,nm ($\HH$-class), 56 $\pm$ 7\,nm ($\MM$-class), and 46 $\pm$ 10\,nm ($\LL$-class) respectively (means $\pm$ SDs, $n=10$ individual tethers). 
The three classes differ from each other with statistical significance: 
$\HH = \MM$ with $P=8.3\times10^{-5}$, $\MM = \LL$ with $P=6.3\times10^{-4}$, and $\HH = \LL$ with $P=1.6\times10^{-5}$ using right-tailed t-tests.

\subsubsection*{A fourth class of Brownian motion was observed in the time-series}
The histograms do not show a second intermediate with a Brownian RMSD different from that of the three classes already mentioned.
Assuming the existence of a second intermediate state $I_2$, this lack of its observation 
suggests that the $I_2$ state is structurally similar to the final-bound state and for that reason does not show up as a separate peak in the histograms of $\B$.
However, there is valuable extra information in the time-series of $\B$ compared to the histograms of $\B$:
The temporal development of $\B$ in Fig.~\ref{Fig:steps_twocol} shows an initial period of multiple transitions to and from the lower class $\LL$ (e.g.,  panel~A, $t\in[100:325]$), followed by a quiescent period with no transitions (panel~A, $t\in[325:650]$).
That is, we observed two classes of $\LL$ with the same Brownian RMSD but with different stability.
We interpret this as the existence of two states of the TBP-DNA complex with equally bent DNA, but with different stabilities.
We identify the first period (fast dynamics) with the second intermediate state $I_2$, and the second period (slow dynamics) with the final-bound state.
The rate constants determined in \cite{Parkhurst1999} indicate that the second intermediate should indeed be populated for only very short periods of time (mean-occupancy-time = $(k_4 + k_5)^{-1}$\,=\,1.2\,s), whereas the final state is populated for a much longer time ($1/k_6$\,=\,282\,s).
Without a second intermediate we cannot explain the observed rapid transitions from the $\LL$ class.

\subsubsection*{Comparison of rate constants to ensemble values}
From manual inspection of ten time-series of $\B$ we estimated the two microscopic rate-constants $k_1$ and $k_3$, and the macroscopic rate-constant $k_a$. 
From the time spent in the $\HH$ class, after addition of TBP and until a transition to the $\MM$ class, we found $k_1 = 1.6 \pm 0.5$\,$\mu$M$^{-1}$s$^{-1}$ (mean $\pm$ SE, $n=10$)\@, in agreement with the ensemble value (1.59\,$\mu$M$^{-1}$s$^{-1}$, \cite{Parkhurst1999}).
From the $\MM$ class we measured transitions to the $\LL$ class and found $k_3 = 54 \pm 16 $\,ms$^{-1}$ (mean $\pm$ SE, $n=11$) again in reasonable agreement with the ensemble value ($\sim$30\,ms$^{-1}$, \cite{Parkhurst1999}). 
Finally, from the waiting time between addition of TBP and until reaching the final-bound state we found $k_a = (9 \pm 3) \times 10^4$\,M$^{-1}$s$^{-1}$ (mean $\pm$ SE, $n=8$), in agreement with reported ensemble values \cite{Parkhurst1999,Powell2001} and with the value we found using silica beads.
The remaining microscopic rate-constants could not be determined with sufficient accuracy, due to the finite time-resolution and sample size, to allow comparison with ensemble data.

\subsection*{What are the Structures of the Intermediates?}
Co-crystal structures of TBP bound to the AdMLP TATA-box show two sharp kinks in the DNA \cite{Kim1993admlp,Kim1994}.  
The first kink occurs where two phenylalanine residues intercalate between the first two base-pairs. The second kink occurs where another pair of phenylalanine residues is inserted between the last two base-pairs (\mbox{5'-T$^{\bigtriangledown}$ATAAAA$^{\bigtriangledown}$G-3'}, triangles indicate intercalations).
Both sets of intercalations produce DNA-kinks of $\sim$45\dg\ \cite{Kim1993admlp,Kim1994}.
Based on the crystal structure, the most straightforward interpretation of our data assigns one pair of intercalations to the $\MM$-class and another pair to the $\LL$-class.
The high flexibility of the 5'-end of the TATA-box implies that this is where the first pair of intercalations is most likely to take place \cite{Juo1996}, whereas the more rigid 3'-end (A-tract) is expected to delay the second set of intercalations.
We therefore propose the following sequence of events: 
Step One, leading to $I_1$, consists of the intercalation of one pair of phenylalanine residues in the  upstream, 5'-end, of the AdMLP\@.
Step Two, leading to $I_2$, consists of the intercalation of the second pair of phenylalanine residues in the downstream, 3'-end, of the AdMLP\@.
Step Three, leading to the final complex, has no major structural change, but consist of a slight rotation of one TBP domain relative to the other \cite{Kim1994}, as well as the formation of van der Waals' contacts between the minor groove of the DNA and the concave surface of the TBP, leading to the stable structure known from crystallography \cite{Kim1993admlp,Kim1994}.  

Assuming a three-step pathway there are two other possible assignments for the two pairs of intercalation events, both of which have previously been proposed:
The first set of intercalations takes place in step Two and the second set in step Three \cite{Parkhurst1999}.
Alternatively, the first set of intercalations take place in step One and the second set in step Three \cite{Powell2001}.
Based on the data presented in this paper we favor the first model over the latter two.
However, we emphasize that it is not possible to make any definitive distinction between these three models based on the existing data---all assignments of structures to states are speculation, so far.

\subsection*{What are the Biological Implications of the Intermediates?}
\begin{figure}[h] 
\begin{center}
	\includegraphics*[width= \linewidth]{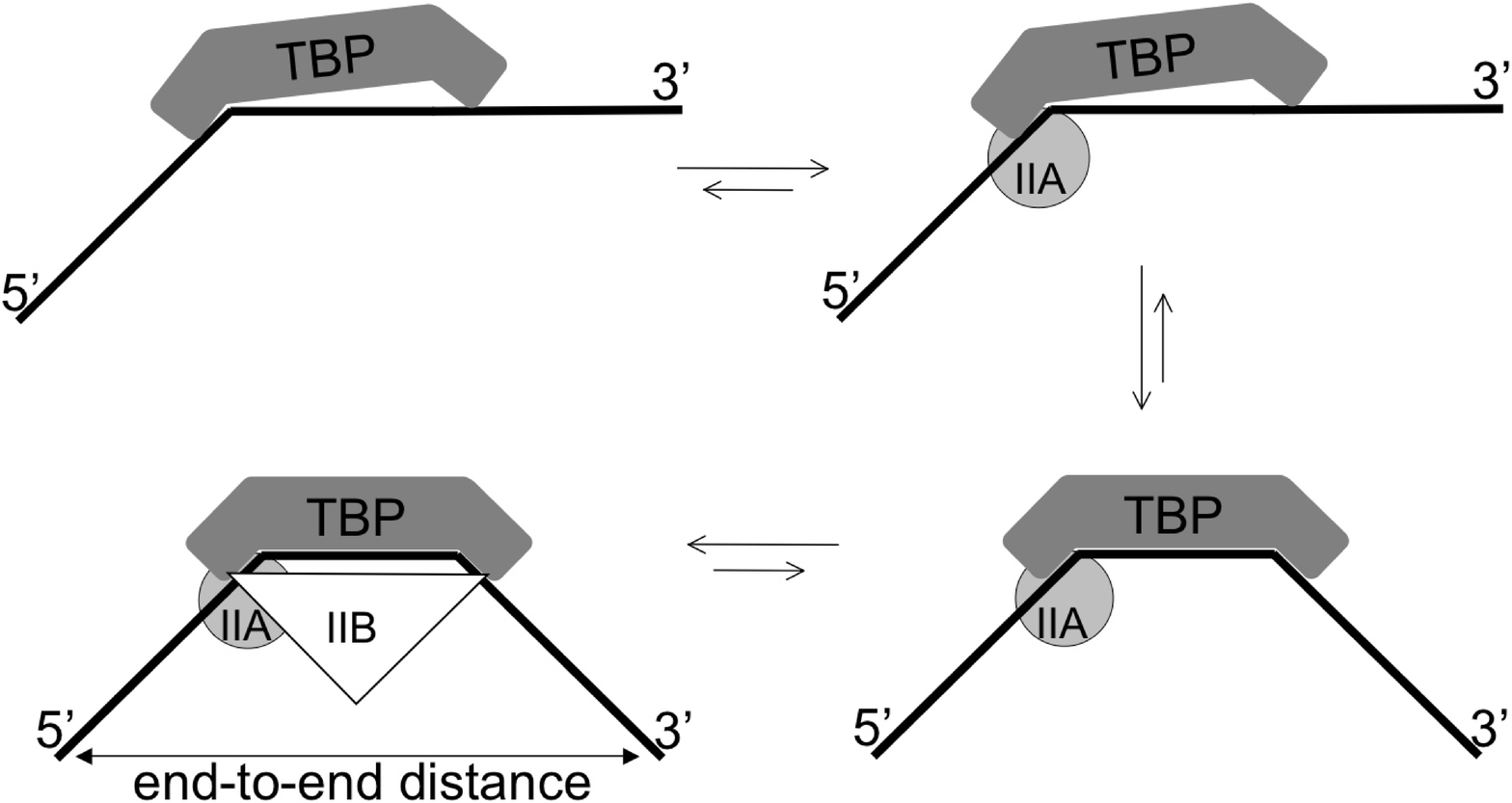}
	\caption{}
	 \label{Fig:TBP_DNA}
\end{center}
\end{figure}

From \textit{in vitro} experiments a picture of the assembly of the transcription pre-initiation complex (PIC) has emerged \cite{Roeder1996}.
In this picture, assembly takes place in steps: TBP binds to DNA, followed by the binding of TFIIB, after which a preformed complex that includes the polymerase is recruited. Later yet, additional factors bind to the complex.  It is possible for TFIIA to enter the PIC at any point after TBP\@.
Now, assume that the $I_1$-state exists and has a single kink in the upstream part of the TATA-box.
This will allow TFIIA to bind stably to the complex, since this factor makes contacts only with upstream DNA-sequences and TBP \cite{Geiger1996,Tan1996}. 
On the other hand, TFIIB makes contacts with both upstream and downstream DNA, as well as the TBP \cite{Nikolov1995}.
In the one-kink $I_1$-state, geometry dictates that proximal to the TBP, upstream and downstream DNA is further apart than in the two-kink $I_2$-state.
In the $I_1$-state therefore,  the formation of a stable association between TFIIB and DNA will presumably be suppressed.
In the $I_2$-state both pairs of phenylalanine intercalations are in place, the DNA is fully bent, and TFIIB can form the contacts known from its crystal structure \cite{Nikolov1995}.  
Thus, we arrive at a picture in which assembly of the PIC can proceed already from the $I_1$-state, and in which the \textit{structural} conformation of $I_1$ suggests an ordering of events with TFIIA binding before TFIIB, see Fig.~\ref{Fig:TBP_DNA} .
The \textit{kinetics} of the $I_1$-state may facilitate the correct orientation of  DNA-bound TBP, as discussed in \cite{Parkhurst1999}.

%%%%%%%%%%%%%%%%%%%%%%%%%%%%%%%%%%%%%%%%%%%%%%%
\section*{CONCLUSIONS}
We made single-molecule experiments investigating the specific binding of a TATA-box Binding Protein to DNA\@. 
In the experiments, beads were attached to a surface by short DNA-tethers and underwent restricted Brownian motion.
When bound by protein the DNA was bent and the Brownian RMSD of the tethered beads decreased.
With this setup we measured kinetic parameters describing the binding of TBP to DNA\@. 

Intriguingly, changes in Brownian motion during a binding event revealed the existence of two intermediate states on the binding pathway. 
This  constitutes the first direct experimental corroboration of a model suggested in \cite{Parkhurst1999}, in which the kinetic pathway of the TBP-DNA interaction has two such  intermediates. 
By direct observation of individual departures and their time of occurrence, we measured kinetic constants describing rates to and from the intermediate states. 
These rates were in agreement with the model-dependent rates derived in \cite{Parkhurst1999} and thus support the kinetic scheme given there.
However, contrary to what is speculated in \cite{Parkhurst1999,Powell2001}, we found that the DNA is less bent in the first intermediate than in the final complex.
This, in turn, might have implications for the order of the assembly of the transcription pre-initiation complex, favoring the association of TFIIA before TFIIB\@.

The results presented here prove that it is possible to make time-resolved observations of single binding and dissociation events of TBP to promoter DNA\@.
This shows that DNA-distortions that are much less dramatic than e.g.\ the looping induced by the \textit{lac} repressor \cite{Finzi1995}, can be reliably detected using well established single-molecule techniques.
Furthermore, the work presented here opens the door to a number of studies of the system, such as quantitative measurements of the force and torque dependence of rate-constants describing the TBP-DNA interaction.

%%%%%%%%%%%%%%%%%%%%%%%%%%%%%%%%%%%%%%%%%%%%%
%%%%%%%%%%%%%%%%%%%%%%%%%%%%%%%%%%%%%%%%%%%%%
\clearpage
\appendix
\section{Appendix}

%%%%%%%%%%%%%%%%%%%%%%%%%%%%%%%%%%%%%%%%
\subsection{Brownian Motion Measure}
\label{app:RMSD}
The time-series of $(x,y)$-positions for the DNA-tethered bead was broken into non-overlapping segments.
Each segment contained a distribution of positions in the $(x,y)$-plane for which we calculated the tensor of inertia.
In two dimensions, the tensor of inertia is a two-by-two matrix
\be
 	\bf{ \tilde{ I } } = \left( \begin{array}{cc} I_{xx}\ & I_{xy} \\ I_{yx} & I_{yy} \end{array} \right)
		\enspace.
\ee
The principal moments are the entries of the diagonalized matrix, and we denote them $I_{\rm min}$ and $I_{\rm max}$ in order of increasing magnitude. 
They are:
\bea
	I_{\rm min} &=& 
	\frac{1}{2}\left\{I_{xx} + I_{yy} - \sqrt{(I_{xx}-I_{yy})^2 + 4I_{xy}^2}\right\} \nonumber\\
	I_{\rm max} &=& 
	\frac{1}{2}\left\{I_{xx} + I_{yy} + \sqrt{(I_{xx}-I_{yy})^2 + 4I_{xy}^2}\right\} 
	\enspace,
\eea
where 
$I_{xx} = \frac{1}{N-1}\sum_{i=1}^N (y_i-\bar{y})^2$, 
$I_{yy} = \frac{1}{N-1}\sum_{i=1}^N (x_i-\bar{x})^2$,   
$I_{xy} = I_{yx} = \frac{1}{N-1}\sum_{i=1}^N (x_i-\bar{x})(y_i-\bar{y})$, 
$x_i$ and $y_i$ are the recorded positions of the tethered bead, 
and $\bar{x}$ and $\bar{y}$ are averages calculated from the $N$ positions in each segment.
The sum of the principal moments equals the sum of the variances along the $x$ and $y$ axes: $I_{\rm min} + I_{\rm max} = \sigma_x^2 + \sigma_y^2$.  
As a measure for the amplitude of the Brownian motion of a tethered bead we used $\B=\sqrt{(I_{\rm min} + I_{\rm max})/2}$, and as a measure for the isotropy of the bead-motion we used the ratio $r = \sqrt{I_{\rm min}/I_{\rm max}}$.
Thus, $\B$ is the RMSD for the positions visited by a tethered bead.
The isotropy-measure was  approximately 50--100\% for a bead with only a single DNA tether.
If $r$ was consistently smaller than $\sim$50\% we interpreted this as a bead tethered by two DNA molecules, a polystyrene link, or some other, non-specific interaction, and discarded the data.  An example of non-isotropic motion is shown in Fig.~\ref{Fig:double_tether_cut}.
%%%%%%%%%%%%%%%%%%%%%%%%%%%%%%%%%%%%%%%%%%%%%%%

%%%%%%%%%%%%%%%%%%%%%%%%%%%%%%%%%%%%%%%%%%%%%%%
\subsection{Multiple Tethers}
\label{app:tethers}
\begin{figure}[h]
\begin{center}
	\includegraphics[width=.5\linewidth]{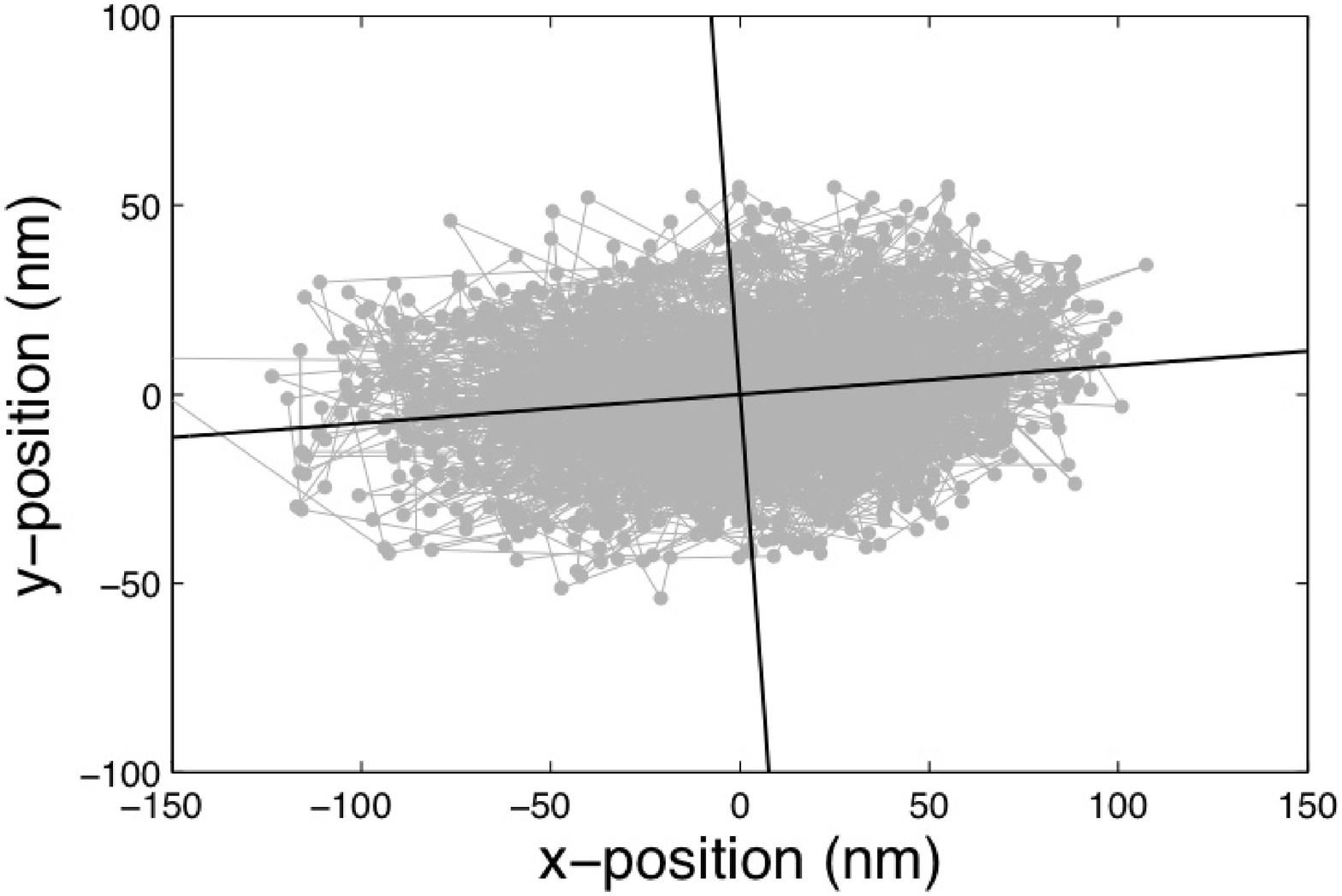}
	\includegraphics[width=.44\linewidth]{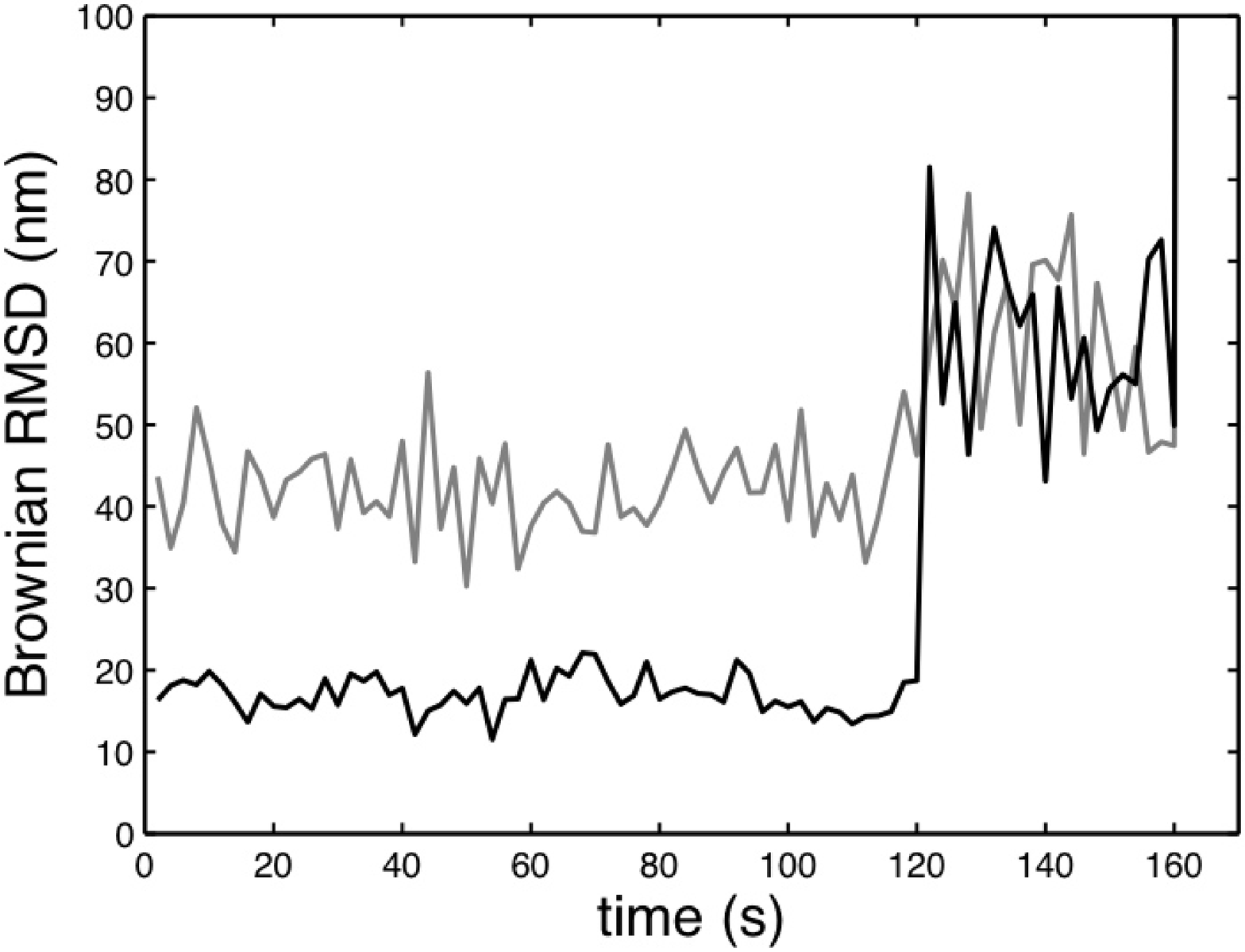}
	\caption{}
	\label{Fig:double_tether_cut}
\end{center}
\end{figure}

We varied the DNA concentration during sample preparation and observed that the fraction of beads with multiple tethers increased as a function of DNA concentration.
Inspection of scatter plots of bead positions revealed several classes of motion: At low DNA concentrations, the vast majority of scatter plots was isotropic; as the DNA concentration was increased some of the scatter-plots were observed to be anisotropic; see Fig.~\ref{Fig:double_tether_cut}, left panel.
At even higher DNA concentrations, approximately isotropic plots with a small radius began to appear along with the previously described shapes.  
These observations are consistent with the formation of one, two, or more tethers per bead: One tether gives rise to an isotropic scatter plot, two tethers yields an elongated, less isotropic, scatter plot, and three or more tethers result in roughly isotropic scatter plots, but with a small radius.
The ease with which multiple tethers are detected owes to the fact that the DNA tether is short, just two persistence lengths.

An example is shown in Fig.~\ref{Fig:double_tether_cut}:
This scatter-plot of bead-positions indicates that two DNA-molecules tethered the bead.
Addition of restriction enzyme lead to a step-wise change in Brownian motion, with
interpretation: Initially the bead was tethered by two DNA-molecules.  
The presence of two tethers broke the rotational symmetry of the setup and forced the bead to move in a quasi one-dimensional fashion.
After 120\,s an enzyme cut one of the DNA tethers and rotational symmetry was restored. 
After an additional 40\,s the other DNA tether was also cut, and the bead diffused away. 
See movie in Supplementary Information.
%%%%%%%%%%%%%%%%%%%%%%%%%%%%%%%%%%%%%%%%5

%%%%%%%%%%%%%%%%%%%%%%%%%%%%%%%%%%%%%%%%%%%%%%
\subsection{Laser Tweezers Stretching Experiments}
\label{app:laser}
Laser tweezers were used to stretch DNA tethers with TBP attached, thereby
forcing TBP off the DNA tether.
For this experiment, flow cells were prepared as described in `Methods', with streptavidin coated silica beads attached to TATA-DNA tethers. 
The protocol included the following steps:
i) TBP was flowed in and allowed to 
bind to the DNA; 
ii) unbound TBP was removed by flowing through 3\,$\times$\,100\,$\mu$l Buffer\,1\@; 
iii) tension was applied to the DNA-tether.
The Brownian motion of the tether was measured before and after each of these steps.
Tension was applied by moving the laser back and forth over the tethered bead six times:
The peak-to-peak distance of the motion was 5\,$\mu$m and the speed was kept constant at 0.1\,$\mu$m/s, i.e., this procedure lasted 5\,min. 
The maximum force exerted on the silica bead by the laser tweezers
was estimated to be 38\,$\pm$\,4\,pN by an escape-method calibration \cite{Svoboda1994}.

\begin{figure}[h]
\begin{center}
	\includegraphics[width=\linewidth]{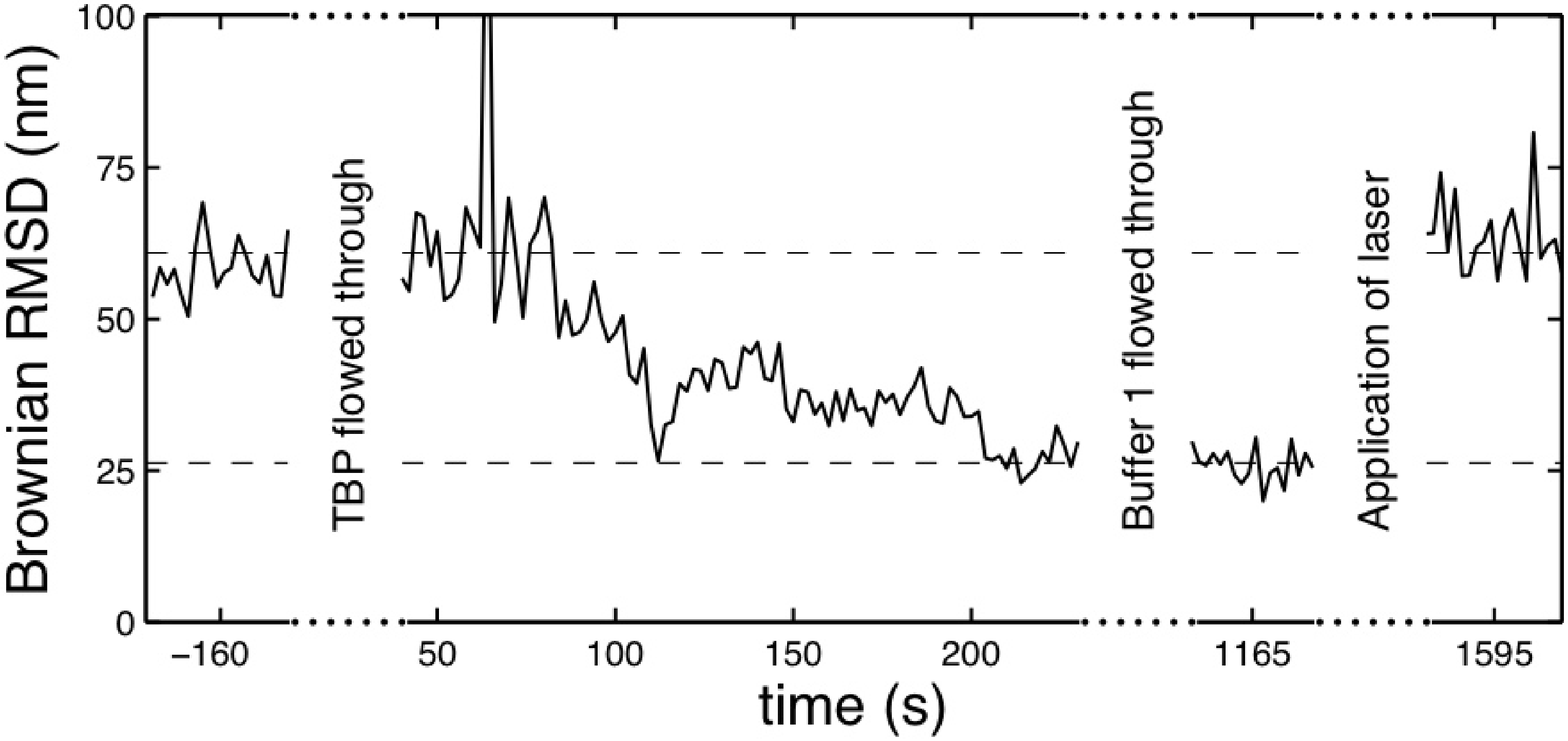}
	\caption{}
	\label{Fig:laser}
\end{center}
\end{figure}

Figure \ref{Fig:laser} shows how the Brownian motion changed during the laser tweezers experiment.
Before TBP was flowed in (at t\,=\,0 seconds), $\B$ was equal to the length of a normal, unbent tether. 
After TBP was flowed in, it bound to DNA.
Unbound TBP was washed out 20\,min after it was flowed in, the Brownian motion was measured, and the laser tweezers were applied for the 5\,min stretching procedure.
After application of the laser tweezers, $\B$ returned to the value of an unbent DNA tether.
We interpret this length-change as the dissociation of TBP from DNA\@. 
Due to irreversible sticking of the bead to the coverslip, the Brownian motion of the tethered bead was determined in only four cases after application of the laser tweezers.
Results similar to those shown in Fig.~\ref{Fig:laser} were found in all cases.
%%%%%%%%%%%%%%%%%%%%%%%%%%%%%%%%%%%%%%%%%%%%%%

%%%%%%%%%%%%%%%%%%%%%%%%%%%%%%%%%%%%%%%%%%%%
\subsection{Weak Entropic Forces Tense the DNA-Tether}
\label{app:entropy}
In  tethered-particle-motion-experiments no external forces are applied.
However, the configuration of the system gives rise to a weak entropic force that tends to stretch the DNA-tether:
If no tether were present, the bead would diffuse away (see movie in supplementary information). Since it does not,  a force must be acting on the bead. The tether is mediating this force, hence is under tension. 
To find the tension in the DNA we first write down the partition function $Z$ for the system, i.e., the sum of all possible configurations in which we can find our system (ignoring all gravitational, inertial, electrostatic, and hydrodynamic effects):
\begin{eqnarray}
  Z(\ell,R) &=& 
  \int_{0}^{\ell} \! g(l) \, \mbox{d}l 
  \int_{0}^{\pi/2} \! l \, \mbox{d}\theta 
  \int_{0}^{2\pi}  \! l \sin\theta \, \mbox{d}\phi \nonumber  \\
  &&\int_{0}^{\alpha_{max}} \!   R \, \mbox{d}\alpha      
  \int_{0}^{2\pi} \!  R \sin\alpha \, \mbox{d}\beta
  \enspace ,
  \label{eq:Z1}  
\end{eqnarray}
where $\alpha_{max} = \cos^{-1} (1-\frac{l}{R} \cos\theta )$,
$g(l) \propto e^{-\beta E(l)}$ is the Boltzmann weight-factor, and $E(l)$ is the energy associated with the tether extension (see Fig.~\ref{Fig:geometry}).  
Integrating over $\alpha$, $\beta$, $\theta$, and $\phi$ we are left with:
\be
  Z(\ell,R) =  2 \pi^2 R   \int_{0}^{\ell} \! l^3 g(l) \, \mbox{d}l 
  \enspace ,
  \label{eq:Z2}
\ee
In the Micro-Canonical Ensemble, the particle number and energy of a system is fixed; in the Canonical Ensemble only the particle number is fixed; in the Grand-Canonical Ensemble both the the particle number and energy can vary. Our system consists of one DNA-molecule and one bead, and this number is fixed. However, the system is in thermal contact with the buffer, so its energy can vary. Thus, our system is described by the Canonical Ensemble and the free energy $H$ of our system is 
\be
  H = -\kB T \ln Z
  \enspace ,
\ee
where $\kB$ is Boltzmann's constant and $T$ is the absolute temperature of the buffer solution.  
From this expression we directly find the tension $F_{\rm DNA}$ by differentiating with respect to $\ell$
\be
  F_{\rm DNA} = -\frac{ \partial H }{ \partial \ell } = \frac{\kB T}{Z} \frac{ \partial Z }{ \partial \ell }
  \enspace .
\ee
If the tether is a stiff rod of length $\ell$, $g(l) \propto \delta(l-\ell)$, where $\delta$ is Dirac's delta function.
In this case the tension is given by the simple expression
\be
  F_{\rm DNA} = 3 \frac{\kB T}{\ell}
  \enspace .
  \label{eq:tension}
\ee

Because work is done against a force when DNA is bent by TBP, the rate of association is expected to be reduced. 
Making the approximation that the association rate-constant decreases as
\be
	k_a = k_0 e^{-F_{\rm DNA} \Delta \ell/\kB T}
	\enspace,
\ee
where $k_0$ is the rate measured in bulk with no external forces applied, we estimate that the association rate should be reduced by 25--50\% due to tension in the DNA-tether \cite{Evans1998}:
In bulk, the end-to-end distance of $\sim$300\,bp dsDNA is approximately 15\% shorter than its contour length \cite{Wilhelm1996}.
Thus, when no external force is applied, the end-to-end distance of a 324\,bp DNA is expected to be 94\,nm, assuming a 0.34\,nm axial rise per bp \cite{Yanagi1991}.
If we model the DNA as a stiff rod of length $\ell = 94$\,nm, an 80$^{\circ}$ kink in the middle of the rod, decreases the end-to-end distance of the rod by $\Delta \ell = 22$\,nm.
The actual change in end-to-end distance is likely to be somewhat smaller due to the flexibility of the DNA and the presence of the force $  F_{\rm DNA}$. A lower limit of $\Delta \ell \ge 7.3$\,nm is set by FRET experiments \cite{Wu2001a}

%%%%%%%%%%%%%%%%%%%%%%%%%%%%%%%%%%%%%%%%%%%%%%

\clearpage
\section*{SUPPLEMENTARY INFORMATION}
An online supplement to this article can be found by visiting BJ Online at http://www.biophysj.org.

\vspace{1cm}
\noindent{\small
\textit{S. cerevisiae} TBP was a generous gift from Michael 
Brenowitz and Elizabeth Jamison. 
We are grateful to Stanley Brown for help and advice on constructing the DNA\@.
LabView software for acquisition of de-interlaced images was kindly provided by Francesco Vanzi.
We thank M. Brenowitz, H. Flyvbjerg, J. Gelles, E.-M. Sch\"otz, and I. M. Toli{\'{c}}-N\o rrelykke for comments on the manuscript.
This work was supported by the Danish research councils and
the Carlsberg Foundation.}

%%%%%%%%%%%%%%%%%%%%%%%%%%%%%%%%%%%%%%%%%%%%%%%
\clearpage
%\bibliography{/Users/simon/00Simon/Project_TATA/bibliography_tata}% .bib fil

%%%%%%%%%%%%%%%%%%%%%%%%%%%%%%%%%%%%%%%%%%%%%%%
%%%%%%%%%%%%%%%%%%%%%%%%%%%%%%%%%%%%%%%%%%%%%%%
\clearpage
\section*{FIGURE LEGENDS}

\subsubsection*{Figure~\ref{Fig:scaledraw}}
Scale-drawing of 0.46\,$\mu$m bead tethered to surface by 324\,bp of DNA, side view.
The full-line circles illustrate the extremal positions the bead can take when the DNA is straight.
The dashed-line circles show the extremal positions the bead can take when the DNA is modeled as two stiff rods at right angles to each other.
The difference between the positions of the center of the bead at the two extrema is nearly 100\,nm.
The Brownian RMSD is a measure for the variance in the bead's $x,y$-positions.
Thus, the change in  Brownian RMSD upon bending of the DNA will be less than the change in extremal positions.

\subsubsection*{Figure~\ref{Fig:silica_event}}
Example of TBP binding to and dissociating from TATA-DNA\@.
\textbf{Main panel:} 
Time-series of the Brownian RMSD of a DNA-tethered 1.0\,$\mu$m silica bead.
Discontinuities of the time-series, due to buffer exchange, are indicated by white spaces on the time-axis. 
Line: Brownian RMSD calculated in non-overlapping 2\,s (50 video-frames) windows.
30\,$\mu$l 226\,nM TBP was flowed through at $t=0$ (first arrow); 
the DNA-tether was bound by TBP after $t=210$\,s;
30\,$\mu$l 1\,M KCl was flowed through at $t=760$\,s;
300\,$\mu$l Buffer~1 was flowed through at $t=850$\,s, around which time the dissociation  took place (second arrow).
\textbf{Inset:} Positions visited by the tethered bead
before (grey, $t\in[112:210]$\,s) and after (black, $t\in[210:270]$\,s) TBP bound to the DNA\@.
The time of binding was determined from the time-series of $\B$ (main panel).

\subsubsection*{Figure~\ref{Fig:kinetics}}
Distribution of waiting times between addition of TBP and observation of a binding event.  
Abscissa: time in seconds.  Ordinate: number of observed events.  
A total of 48 binding events were observed in 29 individual experiments under identical conditions of 226\,nM TBP in Buffer~1 at room-temperature (22.1\,$\pm$\,1.5\,\dg C, mean $\pm$ SD; interval [20\,:\,25]\,\dg C), using TATA-DNA tethered silica beads.
The dynamics of the spatial distribution of TBP in the flow cell was modeled as a diffusive process with reflecting boundary conditions: Setting the diffusion coefficient of TBP to 50\,$\mu$m$^2$/s, the height of the flow cell to 160\,$\mu$m, and the initial distribution of TBP to a delta-function, the distribution of TBP in the flow cell was found to be homogeneous after $\leq$\,60\,s (indicated by vertical dashed line).
We proceeded by excluding from further analysis all events in the first 60\,s  ($n$\,=3).
A  maximum likelihood, single-exponential fit returned a characteristic time of $\tau = 143\pm 22$\,s (mean $\pm$ SE).
White circles show expected counts in each bin, assuming an exponential distribution.
Error-bars shown are expected standard deviations, calculated assuming a binomial distribution of counts in each bin.
The maximum likelihood fit does not depend on the bin-width, because the fit was done directly to the observed waiting times.

\subsubsection*{Figure~\ref{Fig:steps_twocol}}
Direct observation of sub-steps on the TBP-DNA binding-pathway.
Panels A1, B1, and C1 show 3 examples ($n=10$ observed) of the temporal development of the Brownian RMSD $\B$ calculated in non-overlapping 1\,s windows for TATA-DNA tethered 0.46\,$\mu$m polystyrene beads.
100\,$\mu$l TBP was flowed through at time $t=0$ (indicated by arrows). 
Panels A2, B2, and C2 show histograms of $\B$ formed from the data shown in panels A1, B1, and C1, respectively.
Three peaks are present in each of the histograms, corresponding to three classes of Brownian motion.
Horizontal dashed lines in the time-series-panels indicate the positions of the peaks in the histograms.
Multiple back and forth transitions, from the different classes of Brownian motion, can be seen in all three examples.
\textbf{A-panels:} 100\,nM TBP, histogram shows $\B$ in the interval $t\in[50:400]$\,s.
\textbf{B-panels:} 68\,nM TBP, histogram shows $\B$ in the interval $t\in[0:100]$\,s.
\textbf{C-panels:} 68\,nM TBP, histogram shows $\B$ in the interval $t\in[0:290]$\,s. 

\subsubsection*{Figure~\ref{Fig:TBP_DNA}}
Suggested step-wise order of events for the binding of  TBP, TFIIA, and TFIIB to DNA\@.
Starting in the upper left corner, TBP binds to the TATA-box and the first pair of phenylalanines are intercalated in the 5' end of the TATA-box, producing a 45\dg\ kink. This corresponds to the $I_1$ state. Starting from this state TFIIA can bind.  Next, the second pair of phenylalanines are intercalated, in the 3' end of the TATA-box, producing another 45\dg\ kink. This conformation corresponds to the $I_2$ and final-bound state.  In this conformation the DNA is brought close enough together that TFIIB can bind to it.  

%--- figure captions for appendix -------------------------------------
\subsubsection*{Figure~\ref{Fig:double_tether_cut}}
Time series of Brownian motion for a 1\,$\mu$m diameter silica bead tethered by control-DNA, in the presence of \textit{Xho}1.
\textbf{Left panel:} Scatter-plot showing principal axes.
\textbf{Right panel:} The square-root of the two principal moments, $\sqrt{I_{\rm max}}$ and $\sqrt{I_{\rm min}}$, are shown in grey and black, respectively.
During the first 120\,s the motion of the bead was highly anisotropic (isotropy measure $r$\,=39\%).
After 120\,s, the motion of the bead was isotropic ($r$\,=94\%); the bead released from the surface after 160 seconds.

\subsubsection*{Figure~\ref{Fig:laser}}
Time-evolution of Brownian RMSD (calculated in non-overlapping 2\,s windows) during laser tweezers experiment. 
The stepwise binding of TBP to DNA started approximately 80\,s after 30\,$\mu$l 226\,nM TBP was flowed in. 
Approximately 20\,min after addition and binding of TBP, the tweezers were used to pull the silica bead horizontally, thus stretching the tether. 
After this stretching, $\B$ returned to its original value, suggesting that the protein was forced off the DNA\@. 

\begin{figure}[h]
\begin{center}
    \includegraphics[width=\linewidth]{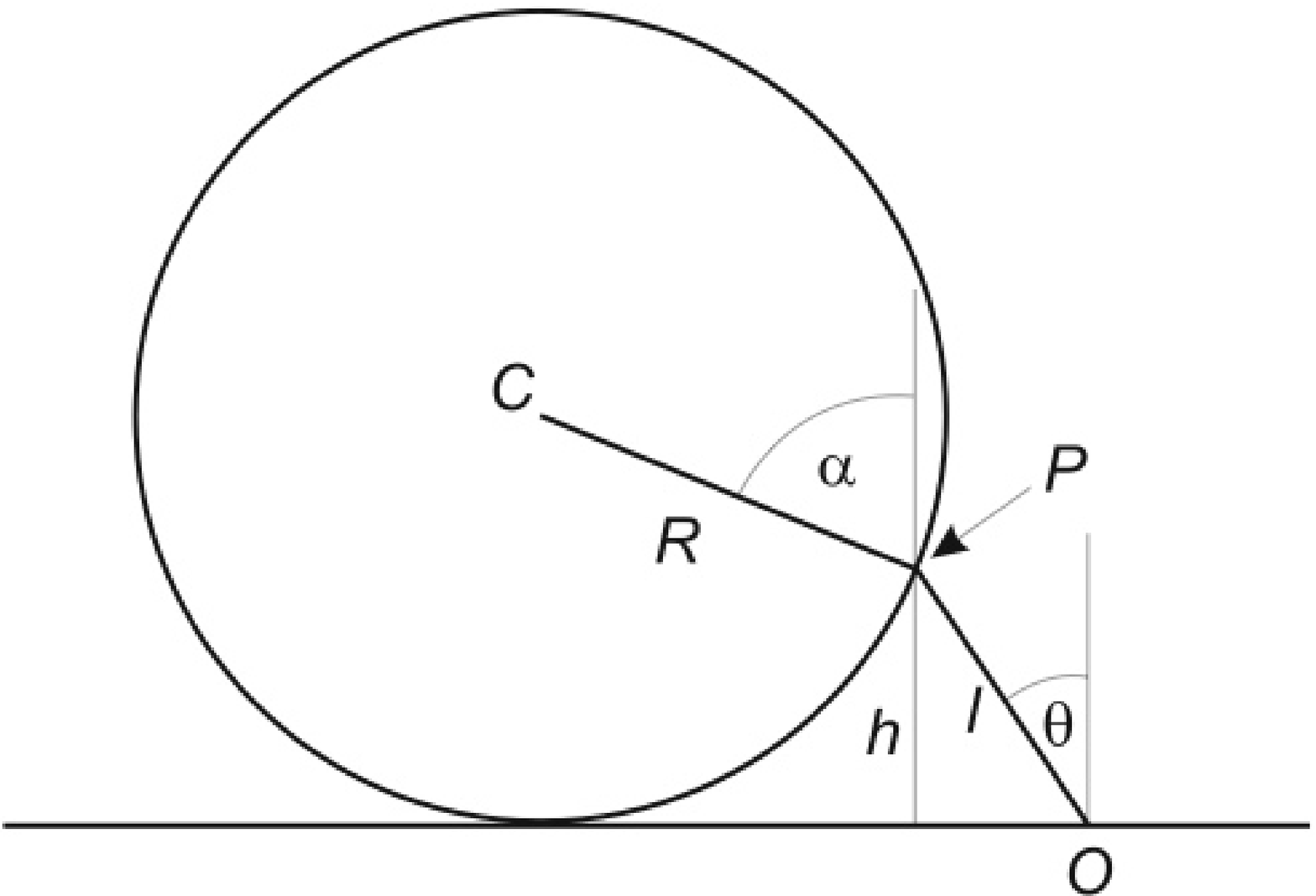}
	\caption{}
	\label{Fig:geometry}
\end{center}
\end{figure}

\subsubsection*{Figure~\ref{Fig:geometry}}
A point $P$ on the surface of a rigid sphere of radius $R$ is attached to a point $O$ on a flat surface by a tether of contour length $L$\@.  
The only constraints on our system are that the sphere stays in the upper half-plane and that the distance $|\vec{OP}| = l \leq L$\@. 
$\theta$ is the angle between $\vec{OP}$ and the surface normal, and $\alpha$ is the angle between $\vec{PC}$ and the surface normal.
Two more angles are needed to fully determine the configuration of the system:  
$\phi$ describes the rotation of $\vec{OP}$ around the surface normal through $O$, and 
$\beta$ describes the rotation of $\vec{PC}$ around the surface normal through $P$\@.
Rotation of the sphere around $\vec{PC}$ contributes an additional degree of freedom, but this degree of freedom is independent of the other parameters, and does not change in our experiment.

%%%%%%%%%%%%%%%%%%%%%%%%%%%%%%%%%%%%%%%%%%%%%%
\onecolumn
\clearpage

%%%%%%%%%%%%%%%%%%%%%%%%%%%%%%%%%%%%%%%%%%%%%%%
\clearpage
%%%%%%%%%%%%%%%%%%%%%%%%%%%%%%%%%%%%%%%%%%%%%%%
\clearpage
%%%%%%%%%%%%%%%%%%%%%%%%%%%%%%%%%%%%%%%%%%%%%%%
\clearpage

%%%%%%%%%%%%%%%%%%%%%%%%%%%%%%%%%%%%%%%%%%%%%%%
\clearpage

%%%%%%%%%%%%%%%%%%%%%%%%%%%%%%%%%%%%%%%%%%%%%%%
%--- Appendix figures --------
%\renewcommand{\thefigure}{\Alph{figure}}
\clearpage

\clearpage

\clearpage

\end{document}